# Superconducting properties and hydrostatic pressure effect on $ABi_3$ (A = Ba, Sr) single crystals

Rajveer Jha, Marcos A. Avila and Raquel A. Ribeiro

CCNH, Universidade Federal do ABC (UFABC), Santo André, SP, 09210-580 Brazil

We report on the crystal growth and characterization of $ABi_3$ (A = Ba, Sr) superconductors. Single crystals of both compounds were grown by the self-flux technique. $BaBi_3$ crystallized in a tetragonal structure with space group *P4/mmm* and $SrBi_3$ in a cubic structure with space group *Pm-3m*. Superconductivity at $T_c$ = 6.0 K for $BaBi_3$ and $T_c$ = 5.6 K for $SrBi_3$ have been confirmed through dc magnetic susceptibility and electrical transport measurements. The dc magnetic susceptibility under hydrostatic pressure shows a positive pressure coefficient of $dT_c/dP$ = 1.22 K/GPa for $BaBi_3$ and a negative pressure coefficient of $dT_c/dP$ = -0.48 K/GPa for $SrBi_3$. The normal state electrical resistivity shows that both compounds are highly metallic in nature. The upper critical fields $H_{c2}$ evaluated by resistivity under magnetic fields $\rho(T,H)$ are 22 kOe for $BaBi_3$ and 2.9 kOe for $SrBi_3$. A specific heat jump of $\Delta C_e/\gamma T_c$ = 1.05 suggests weak coupling superconductivity in $BaBi_3$, whereas $\Delta C_e/\gamma T_c$ = 2.08 for $SrBi_3$ is higher than the BCS theory value of 1.43, indicating a strong coupling superconductor.

**\*Corresponding Authors**
Prof. Dr. Raquel A. Ribeiro
raquel.ribeiro@ufabc.edu.br
and
Dr. Rajveer Jha,
rajveerjha@gmail.com

## Introduction

Bismuth is a very interesting element for condensed matter physics due to its electronic behavior: low carrier density, small effective mass, high mobility, long mean free path, and a large *g*-factor. Because of such interesting electronic properties, bismuth plays an important role in the discovery of new superconducting compounds such as the topological superconductors $Bi_2X_3$ (X=Se, Te), $Bi_4O_4S_3$, $REO_{1-x}F_xBiS_2$ (RE=La, Ce, Nd, and Pr) and others [1-7]. Despite having already been studied in the past, Bi-rich superconductors $ABi_3$ (A = Ba, Sr) have attracted renewed attention in recent times. The older studies on these materials resulted rather limited in depth, probably due to their rather complex synthesis and air sensitivity. However, the superconductivity was reported in 1952 for polycrystalline $ABi_3$ by Matthias and Hulm [8]. Recently, various research groups have reported the superconducting properties of $ABi_3$ for single crystalline compounds [9-11]. In these reports, heat capacity analysis suggested that $BaBi_3$ with superconducting transition $T_c$ = 5.9 K is a weakly coupled superconductor whereas $SrBi_3$ with $T_c$ = 5.75 K is a strongly coupled superconductor. Theoretically, it has been suggested that $BaBi_3$ is a 3D-metal and the Bi $p$-bands are dominant at the Fermi level, resulting in a complex Fermi surface [9]. In addition, a comparatively large mass enhancement has been proposed for $SrBi_3$, in which the Bi-$6p$ electrons are the essential contributors to the Fermi surface [10]. Interestingly, the $T_c$ has been increased to 9.0 K with Na doping in $SrBi_3$ obtained by a high-pressure and low-temperature synthesis technique [12]. The authors interpret the increase as a result of tuning the number of valence electrons, but disregard the effect of chemical pressure. However, external physical pressure ($P$) plays a significant role on the $T_c$ of most metallic and intermetallic superconductors [13-20]. Generally, the applied pressure can change the electronic structure, phonon frequencies and/or electron-phonon coupling, all of which will affect $T_c$ and result in positive or negative pressure coefficients ($dT_c/dP$) [13-20]. Therefore, it was deemed worthwhile to apply hydrostatic pressure on these compounds and follow the effect on their superconducting properties.

In this work, we report on the synthesis and superconducting properties of $ABi_3$ (A = Ba, Sr) single crystals, obtained by the self-flux growth method. $BaBi_3$ crystallized in a tetragonal structure with space group *P4/mmm*, and $SrBi_3$ crystallized in a cubic structure with space group *Pm-3m*. Magnetic susceptibility and electrical transport measurements confirmed that $BaBi_3$ becomes superconducting below 6 K, which is slightly higher than the earlier reported value of 5.6 K [8], and $SrBi_3$ is a superconductor below 5.6 K. Both compounds are highly metallic based on normal state electrical resistivity measurements, and the calculated upper critical fields ($H_{c2}$) through resistivity under magnetic field $\rho(T)|_H$ data are 22 kOe for $BaBi_3$ and 2.9 kOe for $SrBi_3$. Heat capacity measurements for the studied $BaBi_3$ suggests weakly coupled superconductivity, whereas $SrBi_3$ is a strongly coupled superconductor. The dc magnetic susceptibility under hydrostatic pressure for $BaBi_3$ reveals a positive pressure coefficient of $dT_c/dP$ = 1.22 K/GPa, and a negative pressure coefficient $dT_c/dP$ = -0.48 K/GPa for $SrBi_3$. These are the first reports of superconductivity under hydrostatic pressure on ABi$_3$ (A=Ba and Sr) superconducting compounds.

**Experimental Details**

Single crystalline samples of ABi$_3$ (A=Ba, Sr) were prepared through Bi-self flux technique. Ba, Sr pieces (99.95%, Alfa Aesar) and Bi pieces (99.99%, Alfa Aesar) with molar ratio 1:4 were weighed in an Ar-filled glove box then sealed in an evacuated quartz tube. The sealed quartz tubes were slowly heated to 600 °C and dwelled for 10 hours, then slowly cooled down to 350 °C at 2 °C/h. Finally, the quartz tubes were inverted and quickly spun into a centrifuge to remove the excess Bi flux. Rectangular single crystals with shining surfaces and sizes about 1.5×1.5×1.0 mm$^3$ were obtained [see Fig. 1]. The single crystals are air-sensitive and thus they were kept inside the glove box until characterization, to avoid decomposition of the samples. Room temperature X-ray diffraction (XRD) patterns were taken with CuK$_{\alpha 1}$ radiation (λ=1.54 Å) using a STOE STADI-P powder diffractometer in transmission geometry. Heat capacity and electrical transport properties were measured on a Quantum Design physical properties measurement system (PPMS-14T) and magnetic properties measurements were performed on a Quantum Design magnetic property measurement system (MPMS3-7T). The dc magnetic susceptibility under pressure was measured by using a nonmagnetic Cu-Be cylindrical hydrostatic pressure cell in a pressure range up to 1.3 GPa on the MPMS3. We used Daphne oil 7373 as a liquid pressure transmitting medium and a small piece of tin (Sn) was used as a pressure manometer. An uncertainty in pressure of the order P ~ ± 0.05 GPa was determined from the width of the superconducting transition of the Sn manometer.

**Results and discussion**

Figure 2 illustrates Rietveld refined powder x-ray diffraction patterns at room temperature for ABi$_3$ (A=Ba, Sr). We grouped many small and randomly oriented pieces of single crystals to take the XRD data, and recorded the 2$\theta$ sweep in fast mode with a rate of 4°/min to avoid the decomposition of the samples. The BaBi$_3$ crystallized in a tetragonal structure (distorted AuCu$_3$ structure) with space group *P4/mmm*. The refined lattice parameters are $a$ = 5.06(1) Å and c = 5.13(2) Å. In contrast, SrBi$_3$ crystallized in a cubic structure (AuCu$_3$) with space group *Pm-3m* and a refined lattice parameter $a$ = 5.05(3) Å. The resultant lattice parameters are in good agreement with earlier reported results [8-12]. The unit cell for $ABi_3$ is shown in Fig. 2b, resembling a $AuCu_3$ structure with $Bi_6$ octahedra.

Figure 3a shows the temperature dependent dc magnetic susceptibility of $BaBi_3$ under an applied magnetic field $H$ = 10 Oe. The magnetic susceptibility measurements were taken using two protocols: Zero Field Cooled (ZFC) and Field Cooled (FC). A sharp diamagnetic transition at 6 K testifies the high quality of the $BaBi_3$ single crystal. A large diamagnetic signal is observed below 6 K in the ZFC curve. The value of $4\pi\chi$ (ZFC) at 2 K exceeds −1, indicating that the shielding volume fraction at 2 K is essentially 100% and implies bulk superconductivity in the $BaBi_3$ single crystal. The 5-quadrant isothermal magnetization loop ($M \times H$) at 2.1 K for $BaBi_3$ is shown in inset of Fig. 3a. A clear hysteresis is observed, which indicates a typical type-II superconducting behavior. As the magnetic field increases from zero, the magnetization decreases linearly up to $H_{c1}$ = 400 Oe (at 2.1 K). Above $H_{c1}$, the magnetization starts increasing, reaches zero and becomes positive above an applied field of 14 kOe, i.e., the $H_{c2}$ estimated through magnetization is comparably high in this compound. Fig. 3b presents the lower critical fields ($H_{c1}$) as a function of $T/T_c$. For a superconductor the critical fields at absolute zero temperature provide important thermodynamic information. The lower critical field of $BaBi_3$ at absolute zero temperature has been estimated by fitting the equation $H_{c1}(T) = H_{c1}(0)[1 − (T/T_c)^2]$ (solid red line) to the values extracted from the magnetization curves between 2.1 K – 5 K (inset of Fig. 3b). A similar behavior observed for higher temperatures as long as the sample is in the superconducting state, i.e. below 6 K. The estimated critical fields are $H_{c1}(0)$ = 1.71 kOe and $H_{c2}(0)$ = 14.7 kOe. The large $H_{c2}(0)$ implies a small superconducting coherence length of $\xi_0$ = 149 Å, according to $H_{c2}(0) = \Phi_0/2\pi\xi_0^2$, where $\Phi_0$ = 2.0678×10$^9$ Oe.Å$^2$ [21]. The thermodynamic critical field $H_c(0)$ can be calculated as the arithmetic mean of the upper and lower critical fields at absolute zero temperature, i.e. $H_c = (H_{c1}H_{c2})^{1/2}$, and the resulting value is 5.02 kOe. In the Meissner state, using Ginzburg–Landau theory the upper critical field and thermodynamic critical field are related by $H_{c2} = 2^{1/2}\kappa H_c$. The Ginzburg–Landau parameter $\kappa$ thus estimated is 2.07 > 1/2$^{1/2}$, implying type-II superconductivity in $BaBi_3$ [21]. Further, the penetration depth $\lambda(0)$ is calculated from the relation $\lambda(0) = \kappa\xi(0)$, which results in 309 Å.

The $\chi \times T$ curves in ZFC and FC mode under $H = 10$ Oe shows a clear and sharp diamagnetic transition onset around 5.6 K for the SrBi$_3$ single crystal (Fig. 4a). The hysteresis between ZFC and FC regimes indicates SrBi$_3$ is a type II superconductor. Without correcting for demagnetization or sample size effects, we estimate the superconducting volume fraction (ZFC) to be about 70% of perfect diamagnetism, which suggests bulk superconductivity. The Meissner flux expulsion (FC) is about 45% of the diamagnetic flux expulsion, a characteristic of strong flux pinning. The inset of Fig. 4a displays the 5-quadrant $M \times H$ data at $T = 2.1$ K, which suggests weak type II superconducting behavior. Figure 4b shows the estimated $H_{c1}$ for SrBi$_3$, which has been calculated from the magnetic isotherms in the superconducting state, between 2.1 K - 4.5 K (inset of Fig. 4b). The value of magnetization at 2.1 K starts decreasing with increasing magnetic field, reverts the slope at 210 Oe, increases up to 790 Oe, reaches zero and saturates at an applied field of 800 Oe. Similar behaviors are observed for higher temperatures within the superconducting state, i.e., below 5.6 K. The magnetization curves confirm the type-II nature of the superconductivity with $H_{c1} = 210$ Oe and an upper critical field ($H_{c2}$) of about 790 Oe at 2.1 K. The small value of $H_{c2}(0)$ implies large superconducting coherence length of $\xi_0 = 615$ Å, according to $H_{c2}(0) = \Phi_0/2\pi\xi_0^2$ [21]. The calculated $H_c(0)$ at absolute zero temperature, from $H_c = (H_{c1}H_{c2})^{1/2}$ is 410 Oe. According to Ginzburg–Landau theory in the Meissner state, the upper critical field and thermodynamic critical field are related by $H_{c2} = 2^{1/2}\kappa H_c$. The Ginzburg–Landau parameter $\kappa$ thus estimated is $1.38 > 1/2^{1/2}$, implying type-II superconductivity in SrBi$_3$ [21]. The penetration depth $\lambda(0)$ is calculated from the relation $\lambda(0) = \kappa\xi(0)$, resulting in 848 Å.

The temperature dependence of dc magnetization under various hydrostatic pressures up to 0.75 Pa for a BaBi$_3$ single crystal is shown in Fig. 5a. The measurements were carried out in ZFC protocol under $H = 10$ Oe. We define the $T_c$ onset as the beginning of the diamagnetic response, as shown in Fig. 5a. From the $M \times T$ curve at ambient pressure, the $T_c$ for BaBi$_3$ is 6 K. As the pressure increases up to 0.5 GPa, the $T_c$ increases linearly with sharp onset transitions, up to $T_c \sim 6.6$ K. On the other hand, a broadening of the full superconducting transition appears at applied pressure 0.5 GPa, and may be due to a structural phase transition in BaBi$_3$ single crystal compound. The possibility of a pressure-induced structural phase transition should be further investigated with x-ray diffraction measurements under pressure. For the further applied pressure 0.75 GPa, it is observed that the $T_c$ almost saturated at 6.6 K. Figure 5b shows the evolution of $T_c$ against applied pressure. Linear fitting of the data results in a positive pressure coefficient of $dT_c/dP$ = 1.22 K/GPa for the BaBi$_3$ single crystal. The magnitude of this positive pressure coefficient is comparable to that of superconducting compounds such as LuNi$_2$B$_2$C and MgC$_x$Ni$_3$ [14,15].

Figure 6a shows the temperature dependence of dc magnetization under various hydrostatic pressures up 0.81 GPa for a SrBi$_3$ single crystal, measured in ZFC protocol under $H$ = 10 Oe. The $T_c$ onset at ambient pressure for SrBi$_3$ is 5.6 K. As the applied pressure increases the $T_c$ of the compound decreases linearly up to 0.5 GPa with a negative pressure coefficient $dT_c/dP$ = -0.41 K/GPa. Above 0.5 GPa, superconductivity decreases with a larger negative pressure coefficient $dT_c/dP$ = -0.68 K/GPa. The two slopes in decreasing $T_c$ with applied pressure can be clearly seen in Fig. 6b, and the estimated average value of $dT_c/dP$ is -0.48 K/GPa. The superconducting volume fraction increases at the first applied pressure of 0.01 GPa, then decreases consistently for higher applied pressures. The decrease in $T_c$ with applied pressure has been reported earlier for elemental superconductors like Sn, In, Ta, or Hg, and intermetallic superconductors like MgB$_2$ and BiPd [16-20].

Meanwhile, the density of states (DOS) is large enough in ABi$_3$ (Ba, Sr) to produce strong electron-phonon coupling [9-11], the $T_c$ can be expressed by the McMillan formula [22] as $T_c = (\theta_D/1.45)\exp[-1.04(1+\lambda)/\{\lambda-\mu^*(1+0.62\lambda)\}]$, where $\theta_D$ is the Debye temperature, $\mu^*$ is the Coulomb pseudopotential and $\lambda$ is the electron-phonon coupling constant, which can be defined as $\lambda = N(E_F)<I^2>/M<\omega^2>$, where $N(E_F)$ is the DOS at the Fermi level, $<I^2>$ is the average squared electronic matrix element, $M$ is the ionic mass and $<\omega^2>$ is the average square phonon frequency. It is seen from the McMillan relation that the change of $\lambda$ and $\theta_D$ by pressure will define the sign of $dT_c/dP$. It is well known that the applied pressure induces lattice stiffening and usually reduces the $T_c$. On the other hand, the DOS effect can increase or decrease the $T_c$ depending on the increase or decrease of $N(E_F)$ due to applied pressure. It is clear from the McMillan formula that the dependence of $T_c$ on $\theta_D$ is more complex than the change in $\lambda$. In general, the decrease in $T_c$ can be understood as resulting from the $\theta_D$ increasing with applying pressure, which increases the phonon frequency as $<\omega^2> = 0.5\theta_D^2$ and consequently $\lambda$ decreases. Therefore, the positive $dT_c/dP$ for BaBi$_3$ is possibly originated from the increase of $N(E_F)$ and thus by the enhancement of electron-phonon coupling constant $\lambda$. The negative $dT_c/dP$ for SrBi$_3$ is resultant of the decrease of electron-phonon coupling constant $\lambda$ due to increase in $\theta_D$.

The temperature-dependent electrical resistivity from 2–300 K is shown in Fig. 7a for single crystal BaBi$_3$. The sample is metallic with resistivity decreasing linearly above $T$ = 40 K, but decreasing more rapidly for $T$ < 40 K, and showing a sharp superconducting transition at $T_c$ = 6 K. The resistivity in the temperature range of 6-40 K displays Fermi liquid behavior, which allows to fit the resistivity relation $\rho = \rho_0 + AT^2$. The residual resistivity results in $\rho_0$ = 0.15 m$\Omega$.cm and the impurity scattering factor in $A$ = 1.15 m$\Omega$.cm/K$^2$, which suggests that BaBi$_3$ is a weakly correlated system. Inset-I of Fig. 7a shows the resistivity under various magnetic fields in the superconducting transition temperature region. The sharp superconducting transitions are consistent with the increasing magnetic fields $\rho(T)|_H$ measurements and also indicative of good sample quality. The superconducting transition $T_c$ onset and $T_c(\rho=0)$ shift towards lower temperatures with increasing applied magnetic field, as expected for a typical type-II superconductor. The estimated upper critical field is shown in Inset-II of Fig. 7a through analysis of the magnetoresistance data by selecting the 90% normal state resistivity ($\rho_n$) drop as the transition temperature. We estimate the orbital upper critical field, $H_{c2}(0)$, from the Werthamer-Helfand-Hohenberg (WHH) expression, i.e., $H_{c2}(0) = -0.693T_c(dH_{c2}/dT)|_{T=T_c}$ [23]. The slope is used to calculate $H_{c2}(0)$ = 22 kOe by linear fitting of the $H_{c2} \times T$ data. The obtained value of $H_{c2}(0)$ is smaller than the weak coupling Pauli paramagnetic limit $H^{Pauli}$ = 1.82 $T_c$ = 109.2 kOe, which suggests a conventional superconducting mechanism for BaBi$_3$ [25,26].

Figure 7b presents the $\rho \times T$ measurements in the temperature range 2–300 K for a SrBi$_3$ single crystal. It is a very good metal with resistivity decreasing linearly for $T$ > 50 K, then decreases a little more rapidly for T$_c$<$T$<50 K, and shows Fermi liquid behavior fitted to the formula $\rho = \rho_0 + AT^2$. The residual resistivity is $\rho_0$ = 5.3 $\mu\Omega$.cm and the impurity scattering factor is $A$ = 2.7×10$^3$ $\mu\Omega$.cm/K$^2$, indication of electron-electron interactions in SrBi$_3$. The sharp superconducting transition at 5.9 K is clearly shown in $\rho(T)|_H$ curves in the upper inset of Fig 7b. The $T_c$ is suppressed with applied magnetic field, and the suppression rate at very low fields, up to just 3.0 kOe, indicates weak type-II superconductivity in SrBi$_3$. The lower inset of Fig. 7b shows the estimated upper critical field for SrBi$_3$. We take the 90% criteria of normal state resistivity drop point as the transition temperature to calculate the orbital upper critical field, $H_{c2}(0)$, from the WHH equation, i.e., $H_{c2}(0) = -0.693T_c(dH_{c2}/dT)|_{T=T_c}$ [23]. The slope obtained from linear fitting $H_{c2} \times T$ data is used to calculate $H_{c2}(0) = 2.9$ kOe. The value of $H_{c2}(0)$ is much smaller than the weak coupling Pauli paramagnetic limit $H^{Pauli} = 1.82 T_c = 109$ kOe for SrBi$_3$ [24,25].

The specific heat data divided by temperature $(C_p/T) \times T$ data for a BaBi$_3$ single crystal, with and without an applied magnetic field of 50 kOe, is shown in Fig. 8a. The specific heat is sensitive to low temperature electronic excitations and different gap symmetries have different densities of electronic states, i.e., it gives information about the ground state of the system. The lower inset of Fig. 8a shows the heat capacity in applied field 50 kOe, which is best fit to the relation $C_p(T) = \gamma T + \beta T^3 + \delta T^5$ (solid red line). The evaluated specific heat capacity coefficients are $\gamma = 3.2$ mJ/mol·K$^2$, $\beta = 0.155$ mJ/mol-K$^4$ and $\delta = 0.00581$ mJ/mol-K$^6$, which results in a Debye temperature $\theta_D = (234zR/\beta)^{1/3} = 171$ K. Here $z = 4$ is the number of atoms in the BaBi$_3$ unit cell, and the Rydberg constant $R = 8.314$ J/mol·K is the tern $\gamma T$ is the normal state electronic contribution and $\beta T^3$ and $\delta T^5$ are the lattice contributions to the specific heat. The upper inset of Fig. 8a shows the electronic contribution of specific heat $C_e/T$ as a function of temperature. Here $C_e(T) = C_p - \beta T^3 - \delta T^5$, i.e., the lattice contribution subtracted from the total $C_p$. The observed specific heat jump $\Delta C_e/\gamma T_c = 1.05$ is lower than the BCS theory value of 1.43, suggesting weak coupling superconductivity in BaBi$_3$. Our value is very similar to that obtained by Haldolaarachchige *et al.* [9] for their BaBi$_3$ single crystalline sample.

Figure 8b presents the temperature-dependent specific heat $C_p/T$ for the SrBi$_3$ single crystal, with and without an applied magnetic field of 50 kOe. The specific heat capacity coefficients are calculated as $\gamma = 6.5$ mJ/mol·K$^2$, $\beta = 0.124$ mJ/mol·K$^4$ and $\delta = 0.00581$ mJ/mol·K$^6$, through the relation $C_p(T) = \gamma T + \beta T^3 + \delta T^5$ (solid red line) by fitting the heat capacity data in applied field of 50 kOe seen in the lower inset. The Debye temperature is $\theta_D = (234zR/\beta)^{1/3} = 180$ K. The upper inset of Fig. 8b shows the electronic contribution of specific heat $C_e/T$ as a function of temperature and the observed specific heat jump $\Delta C_e/\gamma T_c = 2.08$ is higher than the BCS theory value of 1.43, suggesting SrBi$_3$ as a strong coupled superconductor. Our result is in reasonably good agreement with the previous reported values of 2.7 by Kempf *et al*. and 2.17 by Shao *et al*. [11, 26].

In summary, the single crystalline BaBi$_3$ and SrBi$_3$ compounds studied in this work have shown different crystal structures, respectively tetragonal with space group *P4/mmm* and cubic with space group *Pm-3m*. The difference in lattice parameters $a$ = 5.06(1) Å and c = 5.13(2) Å is relatively small, i.e., BaBi$_3$ crystallizes in a distorted AuCu$_3$ type structure, while SrBi$_3$ crystallizes in the AuCu$_3$ structure with lattice parameter $a$ = 5.05(3) Å. Both compounds have quite different values of superconducting parameters in the scenario of conventional type-II superconductivity, as summarized in Table 1. The hydrostatic pressure studies for BaBi$_3$ and SrBi$_3$ single crystals show a positive pressure coefficient of *dT$_c$/dP* = 1.22 K/GPa and a negative pressure coefficient *dT$_c$/dP* = -0.48 K/GPa, respectively. Heat capacity studies suggest that BaBi$_3$ is a weak coupled superconductor, whereas SrBi$_3$ is a strongly coupled superconductor.

Acknowledgements

The authors thank P.G. Pagliuso for access to the GPOMS PPMS used for the resistivity measurements. We are thankful to Dr. V. P. S. Awana (CSIR-National Physical Laboratory, New Delhi, India) for his valuable discussions and for reviewing the manuscript. This work is supported by Brazilian agencies FINEP, CNPq and FAPESP (Grant Nos. 2011/19924-2, 2015/15403-9).

Table 1. Summary of the superconducting parameters of $ABi_3$ (A=Ba, Sr) single crystals.

| Parameters | Units | $BaBi_3$ | $SrBi_3$ |
|---|---|---|---|
| $T_c$ | K | 6.0 | 5.6 |
| $H_{c1}(0)$ | Oe | 1710 | 210 |
| $H_{c2}(0)$ | Oe | 14710 | 790 |
| $H_c(0)$ | Oe | 5015 | 410 |
| $\xi$ | Å | 149 | 615 |
| $\lambda$ | Å | 309 | 848 |
| $\kappa$ | Å | 2.07 | 1.38 |
| $dTc/dP$ | K/GPa | 1.22 | -0.48 |
| $\Delta C_e/\gamma T_c$ | | 1.05 | 2.08 |
| $\Theta_D$ | K | 171 | 180 |

**Figure captions**

**Figure 1:** Photos of the as-grown single crystals of (a) BaBi$_3$ and (b) SrBi$_3$.

**Figure 2:** (a) Room temperature X-ray diffraction patterns of ABi$_3$ (A=Ba,Sr) single crystalline clusters (red open circles), Rietveld refinements (black solid line) and differences (blue solid line). (b) Schematic representation of the crystal structure of ABi$_3$ (A=Ba, Sr).

**Figure 3:** (a) Temperature dependence of the dc magnetic susceptibility in ZFC and FC modes for BaBi$_3$ single crystal. The inset is the isothermal magnetization loop at 2.1 K. (b) Lower critical field $H_{c1}$ as a function of normalized temperature $T/T_c$. The red solid line represents the fitting of the equation $H_{c1}(T) = H_{c1}(0)[1−(T/T_c)^2]$. The inset shows the initial (ZFC) isothermal magnetization branches at different temperatures.

**Figure 4:** (a) Temperature dependence of the dc magnetic susceptibility in ZFC and FC modes for SrBi$_3$ single crystal. The inset is the isothermal magnetization loop at 2.1 K. (b) Lower critical field $H_{c1}$ as a function of normalized temperature $T/T_c$. The red solid line represents the fitting of the equation $H_{c1}(T) = H_{c1}(0)[1−(T/T_c)^2]$. The inset shows the initial (ZFC) isothermal magnetization branches at different temperatures.

**Figure 5:** (a) dc magnetic susceptibility as a function of temperature in ZFC mode for a BaBi$_3$ single crystal under various hydrostatic pressures (0-0.75 GPa). (b) $T_c^{onset}$ vs. applied pressure for the BaBi$_3$ single crystal.

**Figure 6:** (a) dc magnetic susceptibility as a function of temperature in ZFC mode for a SrBi$_3$ single crystal under various hydrostatic pressures (0-0.81 GPa). (b) $T_c^{onset}$ vs. applied pressure for the SrBi$_3$ single crystal.

**Figure 7:** (a) Temperature dependence of the electrical resistivity of a BaBi$_3$ single crystal, including a low-$T$ curve fitting of the relation $\rho = \rho_0 + AT^2$. Inset-I shows $\rho(T)$ under various applied magnetic fields. Inset-II shows the upper critical field ($H_{c2}$) as a function of temperature and the red solid line is data fit of the WHH expression (see text). (b) Temperature dependence of the electrical resistivity of a SrBi$_3$ single crystal, including a low-$T$ curve fitting of the relation $\rho = \rho_0 + AT^2$. Inset-I shows $\rho(T)$ under various applied magnetic fields. Inset-II shows the upper critical field ($H_{c2}$) as a function of temperature and the red solid line is data fit of the WHH expression.

**Figure 8:** (a) Heat capacity $C_p/T$ vs. $T$ of a BaBi$_3$ single crystal in the temperature range 4-8 K. The lower inset shows heat capacity under applied field of 50 kOe fitted by $C_p(T) = \gamma T + \beta T^3 + \delta T^5$. The upper inset shows the electronic part $C_e/T$ vs. $T$ around the transition. (b) Heat capacity $C_p/T$ vs. $T$ of a SrBi$_3$ single crystal in the temperature range 4-8 K. The lower inset shows heat capacity under applied field of 50 kOe fitted by $C_p(T) = \gamma T + \beta T^3 + \delta T^5$. The upper inset shows the electronic part $C_e/T$ vs. $T$ around the transition.

Figure 1.

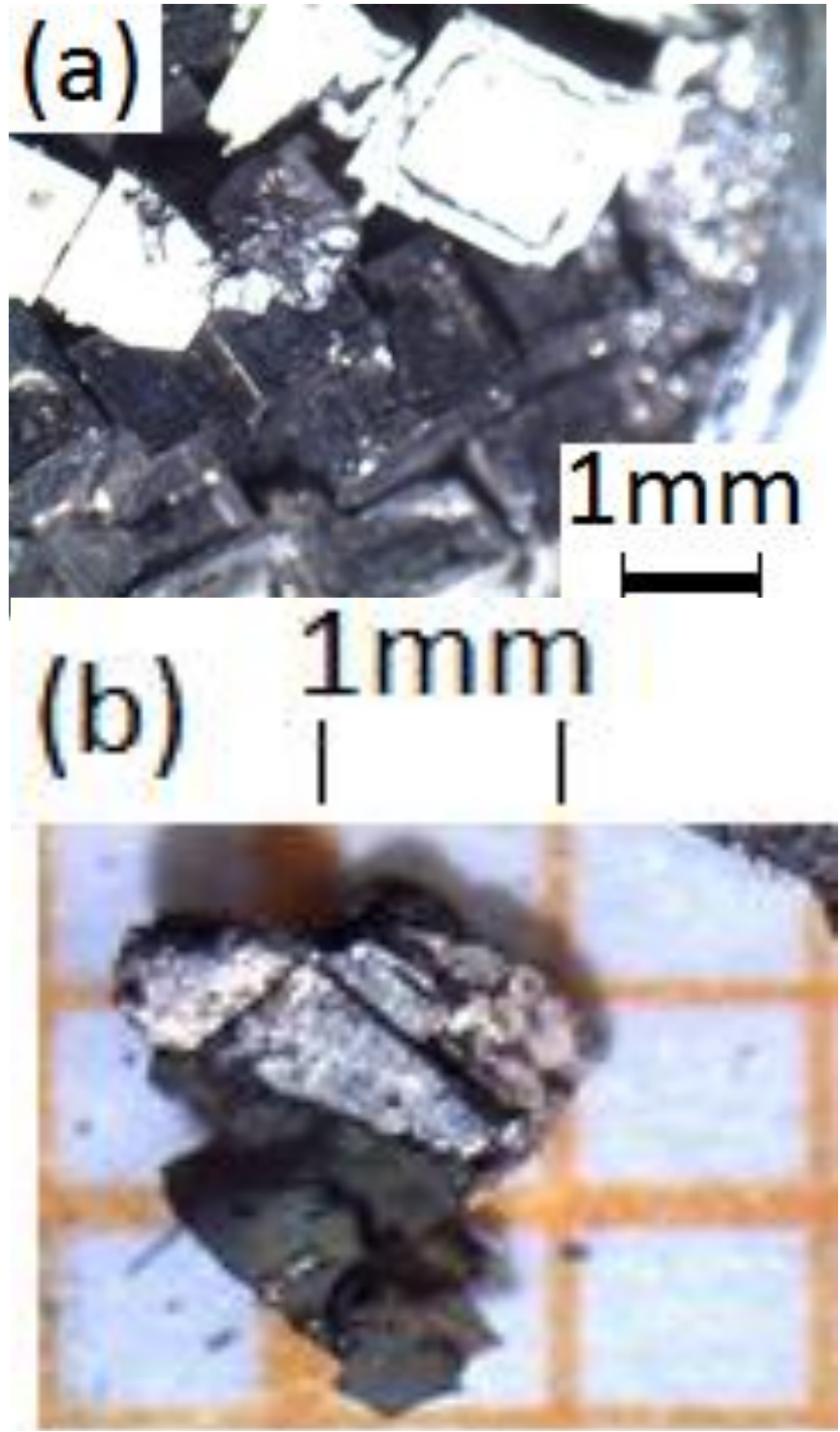

Figure 2.

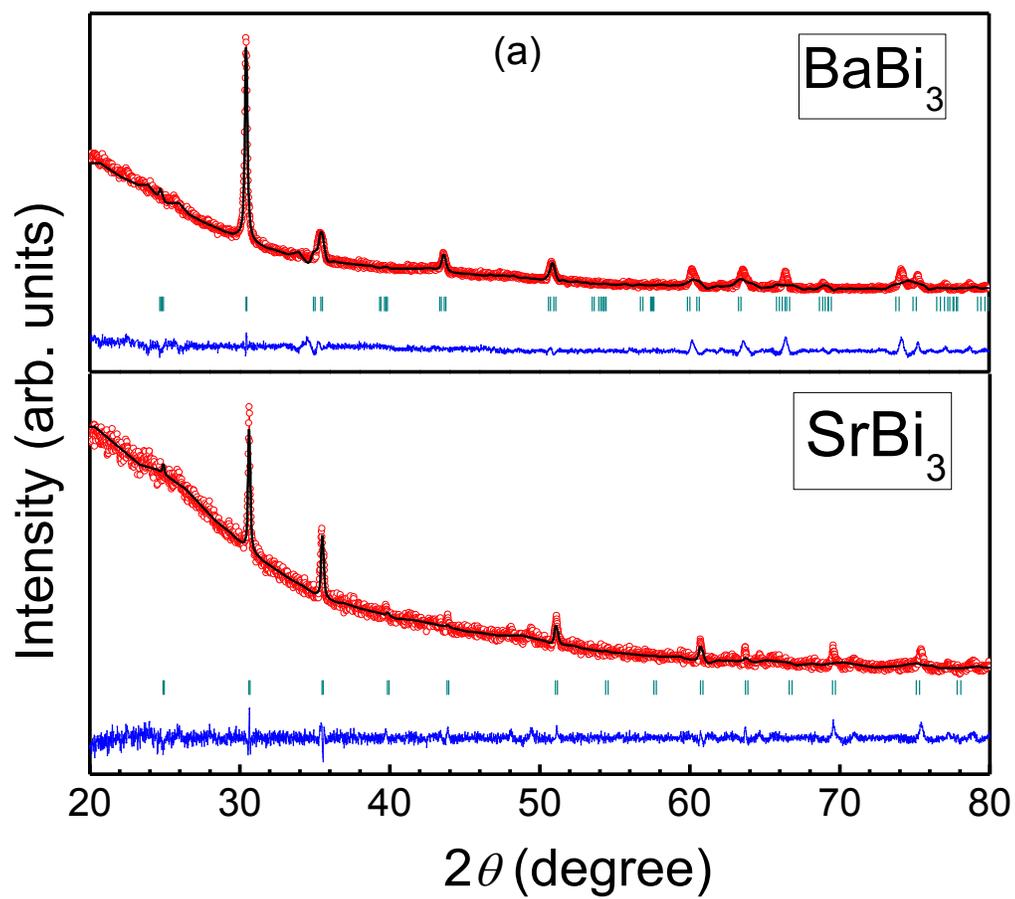

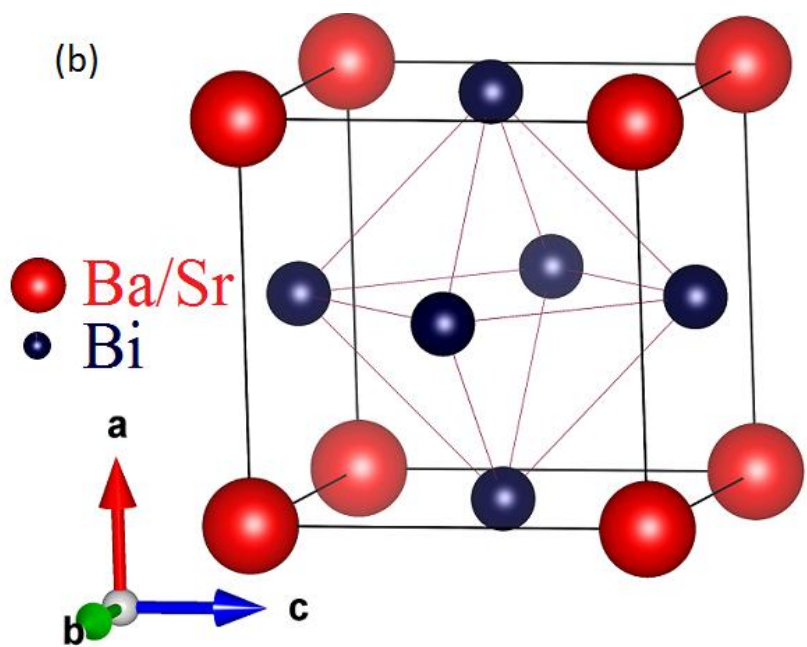

Figure 3.

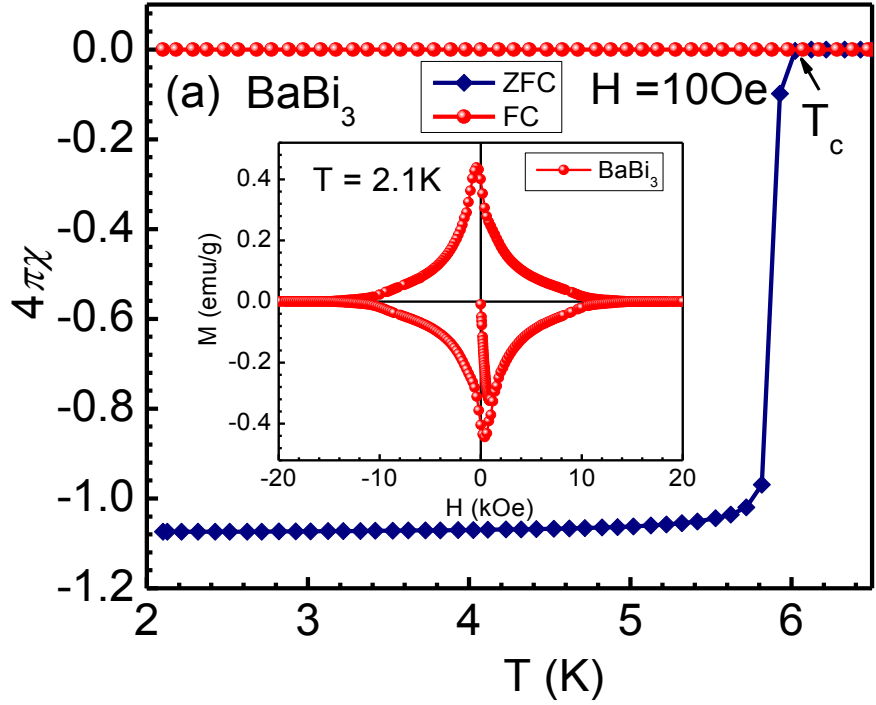

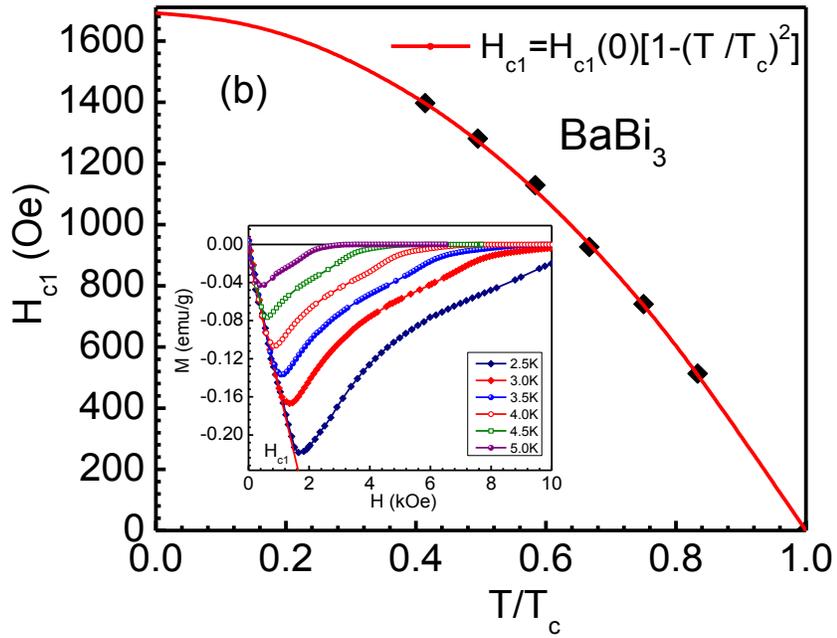

Figure 4.

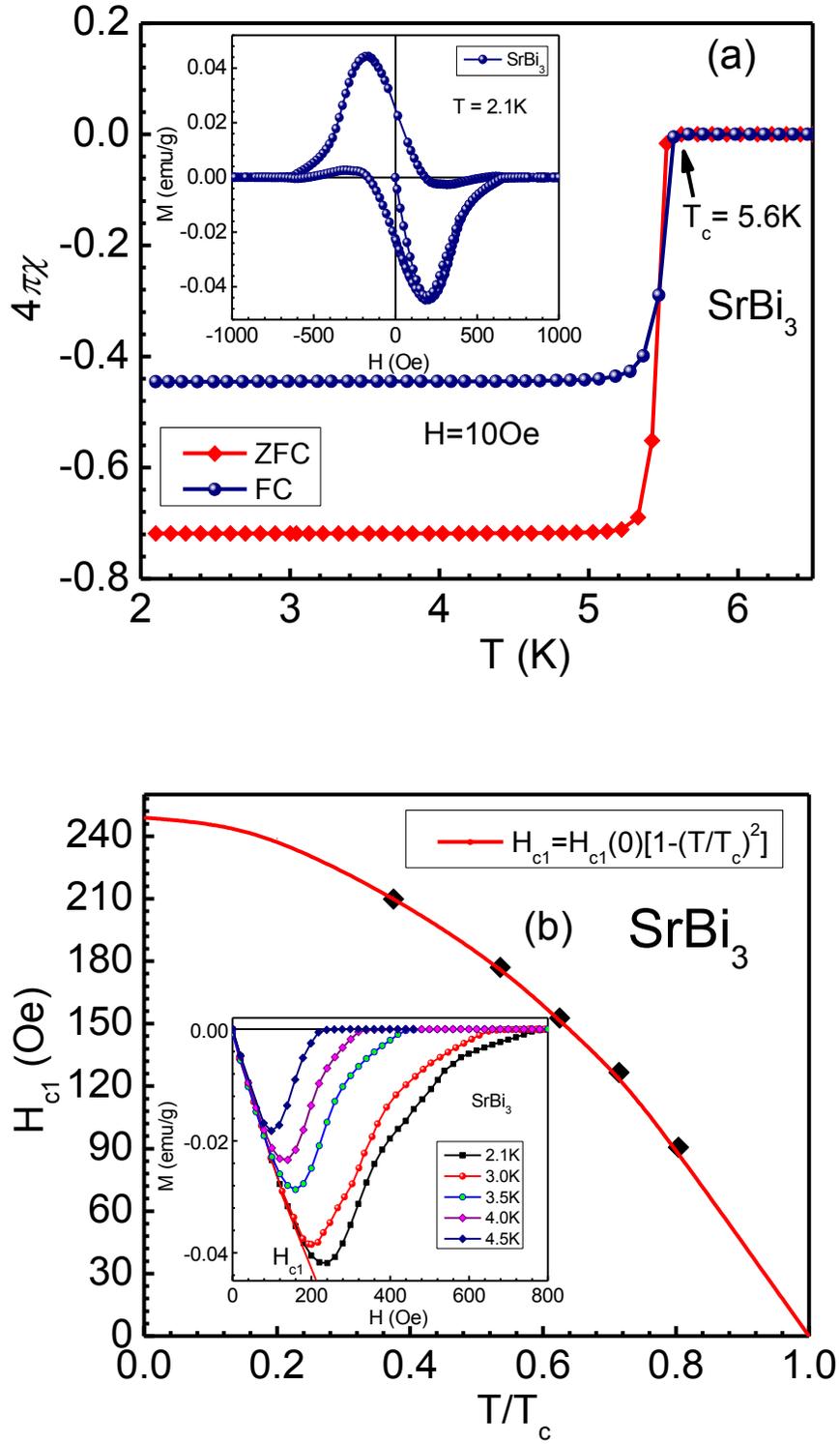

Figure 5.

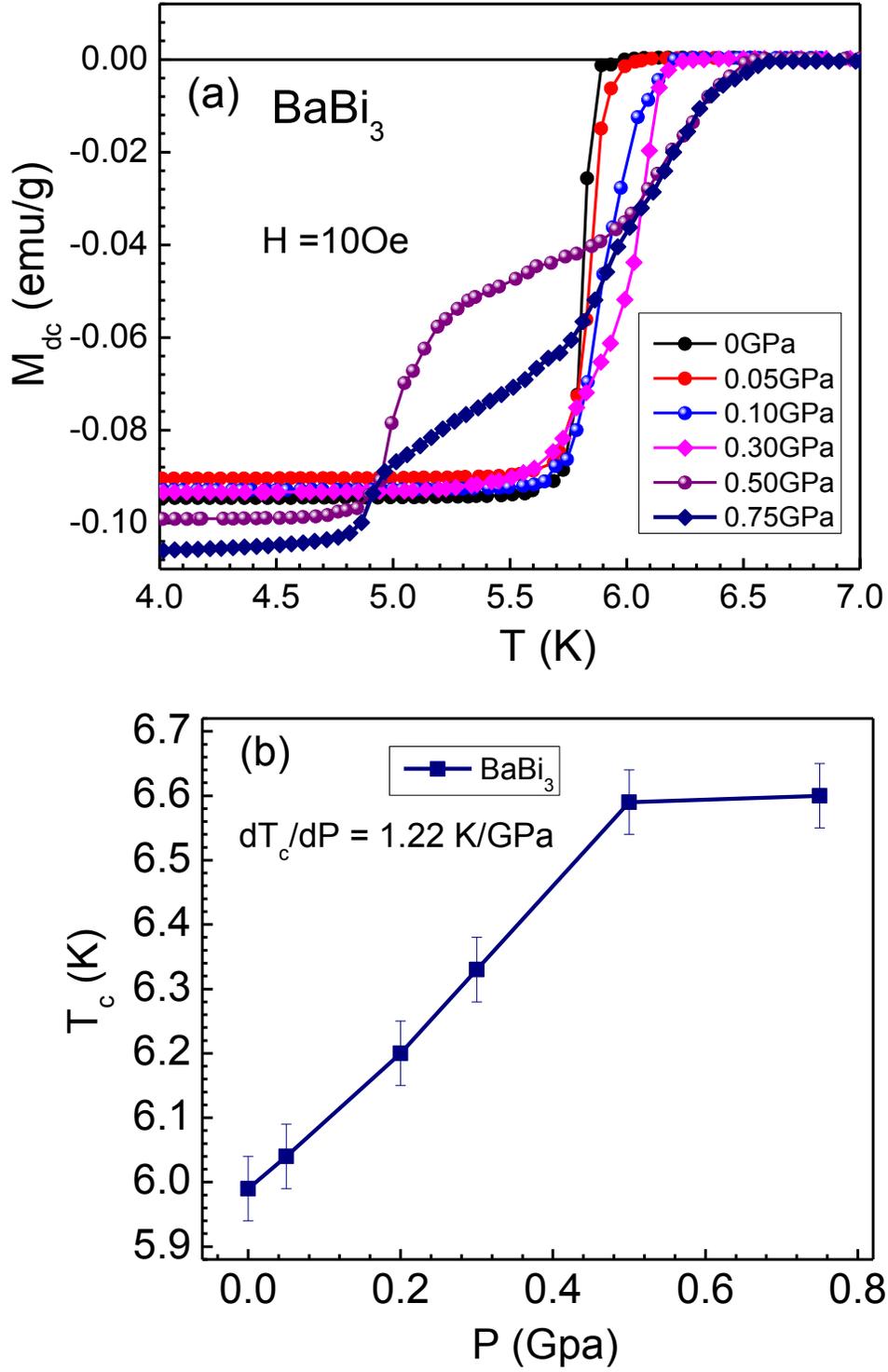

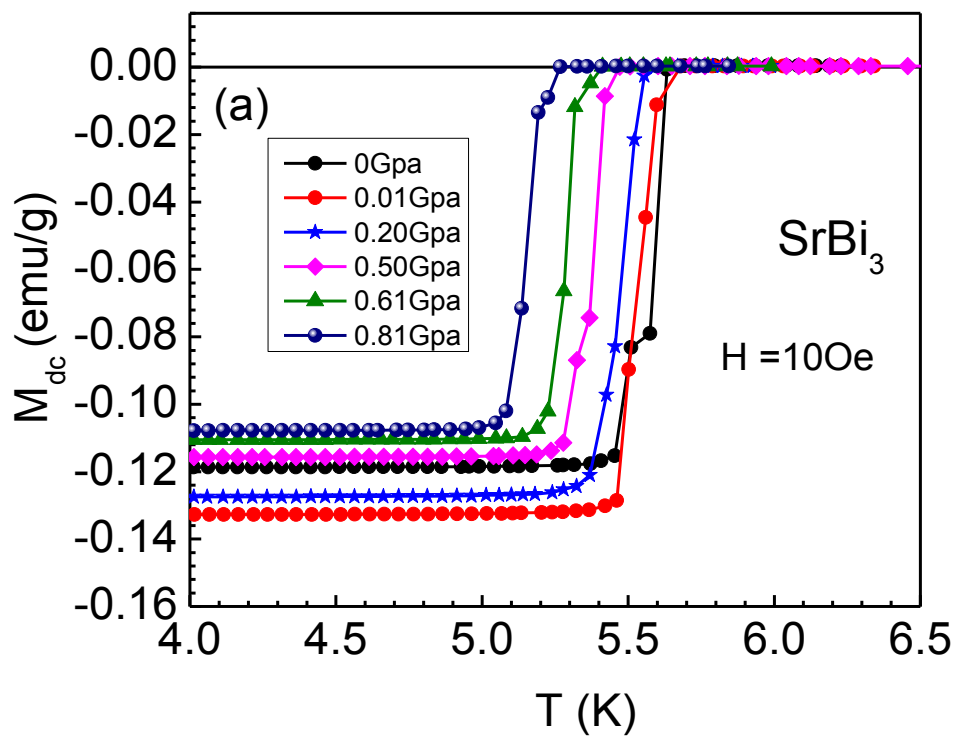

Figure 6.

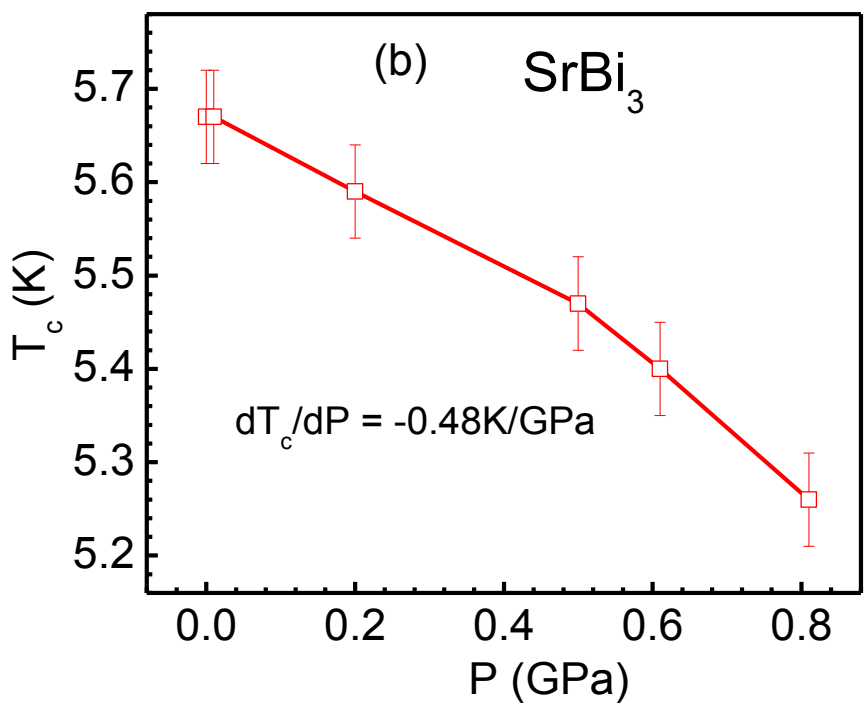

Figure 7.

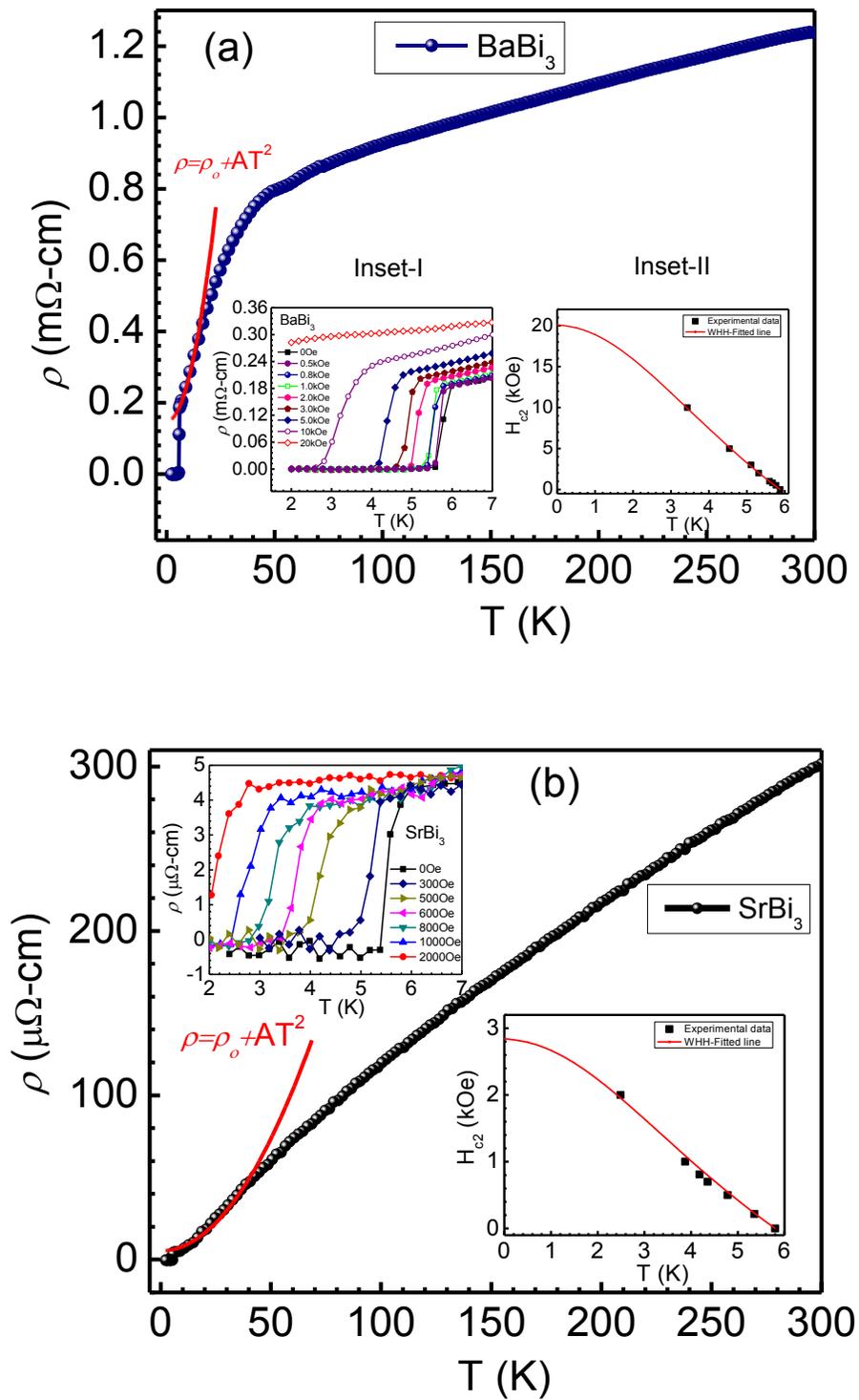

Figure 8.

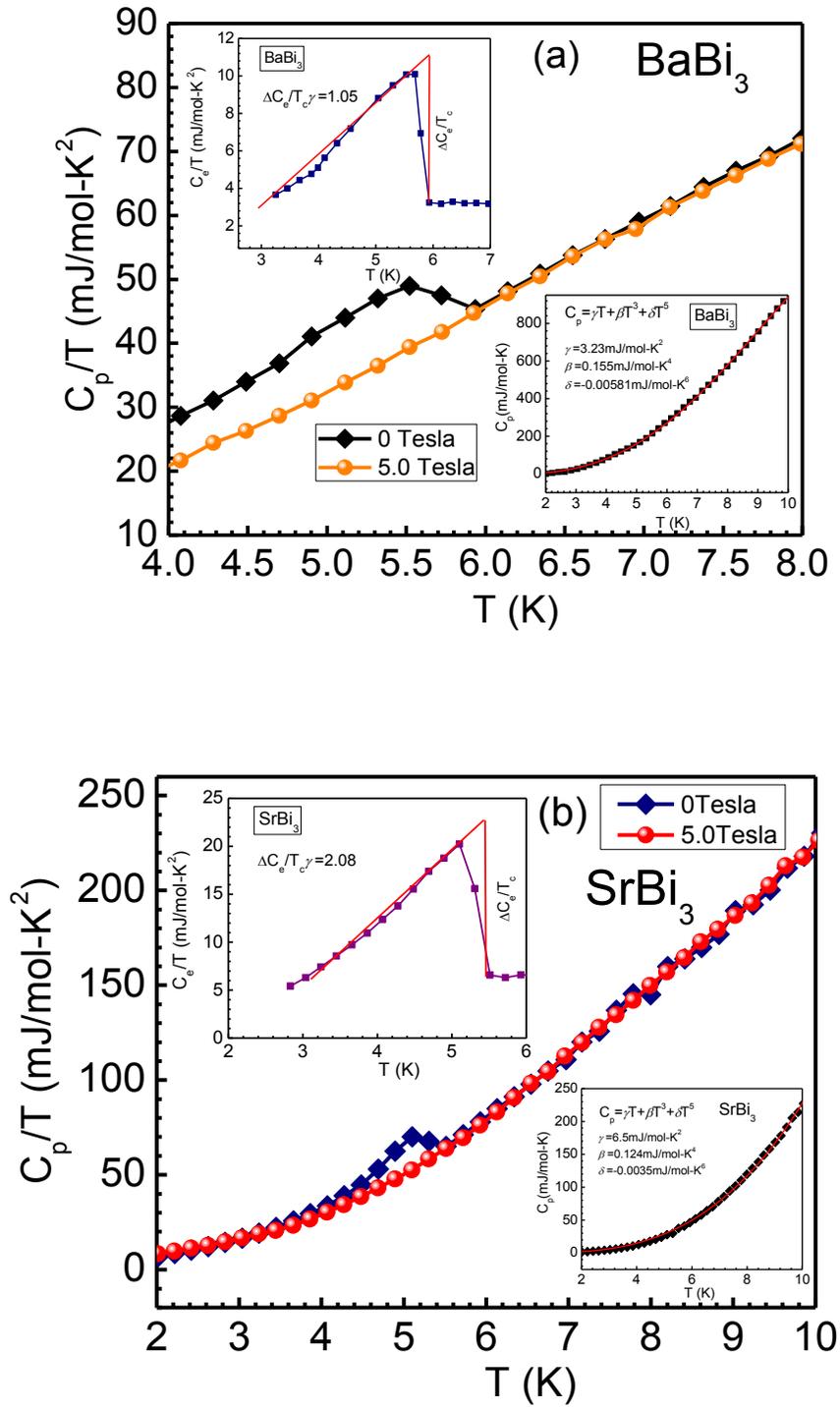